\documentclass[aps,prd,10pt,twocolumn,nofootinbib]{revtex4}

\usepackage{epsfig}
\usepackage{bm}
\usepackage{latexsym}
\usepackage{natbib}
\usepackage{url}
\usepackage{dcolumn}
\usepackage{color}
\usepackage{amsfonts,amssymb,amsmath}
\usepackage{graphicx,epsfig}
\usepackage{psfrag}
\usepackage{subfigure}
\usepackage{hyperref}
\hypersetup{colorlinks=true}
\usepackage{mathtools}
\usepackage{enumitem}
\usepackage{float}
\usepackage{mathrsfs}
\usepackage[dvipsnames]{xcolor}

\begin{document}

\title{Extracting the effective contact rate of COVID-19 pandemic}

\author{
Gaurav Goswami$^{a,}$\footnote{gaurav.goswami@ahduni.edu.in},
Jayanti Prasad$^{b,}$\footnote{prasad.jayanti@gmail.com}\footnote{Currently working as an independent researcher and data scientist.} and
Mansi Dhuria$^{c,}$\footnote{mansidhuria@iitram.ac.in}
}

\affiliation{
$^a$ School of Engineering and Applied Science, Ahmedabad University, Ahmedabad 380009, India\\
$^b$ Khagol-20, 38/1 Panchvati, Pashan, Pune - 411008, INDIA \\
$^c$ Institute of Infrastructure, Technology, Research and Management, Ahmedabad 380026, India
}

\begin{abstract}
\noindent In the absence of any available vaccines or drugs, prevention of the spread of Coronavirus Disease 2019 (COVID-19) pandemic is being achieved by putting many mitigation measures in place.
It is indispensable to have robust and reliable ways of evaluating the effectiveness of these measures.
In this work, we assume that, at a very coarse-grained level of description, the overall effect of all the mitigation measures is that we can still describe the spread of the pandemic using the most basic {\it Susceptible-Exposed-Infectious-Removed} ($SEIR$) model but with an ``effective contact rate" ($\beta$) which is time-dependent. 
We then use the time series data of the number of infected individuals in the population to extract the instantaneous effective contact rate which is the result of various social interventions put in place. 
This approach has the potential to be significantly useful while evaluating the impact of mitigation measures on the spread of COVID-19 in near future. 
\end{abstract}

\maketitle

\section{Introduction}

Within a few months of its first outbreak, Coronavirus Disease 2019 (COVID-19) has infected millions of individuals worldwide and hence, has been recognised as a pandemic by the World Health Organisation \cite{WHO}. We are just beginning to discover various relevant details about this pandemic and severe acute respiratory syndrome coronavirus 2 (SARS-CoV-2), the virus strain that causes it \cite{Gorbalenya-2020,Wu-2020,Zhou-2020}. As of now, no candidate vaccine or drug has successfully completed any clinical trials.
Given this, various preventive and mitigative measures are being practised worldwide
\cite{Ferguson-2006,Huaiyu-2020}.
It is extremely important to evaluate the impact and efficacy of these mitigation measures. 

The most obvious way to answer these questions is to build elaborate mathematical models simulating the spread of the pandemic \cite{Imperial-college,time-dependent-beta,Singh-2020,Prem-2020,Berger,jia-2020,pandey-2020,Pribylova-2020,Das-2020,Castilho-2020,Sardar-2020}.
The speed with which an epidemic spreads is dependent on a large number of factors. E.g., an infectious disease has certain intrinsic parameters such as mean serial interval, mean incubation period, mean infectious period etc. In addition, the spread of the disease also depends on the ease with which newer susceptible individuals can get infected: this plays a pivotal role in determining parameters such as the basic reproduction number of the epidemic in a population. 
The mitigation measures essentially try to ensure that the number of contacts between infectious and susceptible individuals as well as the probability of infection on contact are as small as possible. 
Modelling the spread of an epidemic while the various mitigation measures are being practised is thus a complex task.

Given this, one might ask, {\it if there is a resurgence of the pandemic, could we use the information we gain during the first wave, to build better models which deal with the spread of the pandemic while the myriad mitigation measures are taking place}?
The present work deals with an approach which could prove to be useful when we wish to have a coarse grained description which still takes into account all the different effects (such as migrations, contact tracing, quarantines, lockdowns, heterogeneous mixing of population, testing of some fraction of asymptomatic cases etc) which go on almost simultaneously when an epidemic spreads. We surmise that, as these myriad complex processes take place, at a very broad, coarse grained level of description, the dynamics of a pandemic such as COVID-19, which is known to have some latency period, is still going to be described by the well known Susceptible-Exposed-Infectious-Removed, SEIR model \cite{Book_2008,Book_2018} with an effective contact rate $\beta$ which will be time-dependent.  
We shall present a step-by-step procedure by which this time dependent effective contact rate $\beta (t)$ can be reliably reconstructed from the time series data of infectious fraction of the population.

We must emphasise that this effective instantaneous contact rate $\beta (t)$ is found while all the complicated mitigation measures and social distancing measures are being practised and put in place. Thus, for a given population such as a given city, we shall know the mitigation measures taken, and we can use the procedure described in this work to determine the effective instantaneous contact rate (or one of its smoothed incarnations). 

Since the knowledge of instantaneous effective contact rate lets us reliably calculate the evolution of an epidemic, once this procedure is followed for a large number of localities (with known mitigation measures), one becomes better equipped in handling the impact of various mitigation measures: this can prove to be very useful in planning for future outbreaks.

This paper is organised as follows: in the next section, we review some fundamentals of SEIR class of epidemiological models and describe how we can connect theory to observations. In  \textsection \ref{sec:strategy}, we provide a step-by-step procedure for reconstructing a raw effective contact rate and analyse the robustness of this procedure. In addition, in this section, we present ways of smoothening the raw contact rate obtained and mention some applications, in particular, to the time series data of confirmed cases of many countries. Finally, we conclude with discussions about the possible significance of our approach.

\section{Epidemiological models and observations}

\subsection{SEIR models: a quick reminder}
\label{sec:reminder}

There exists a very large class of models which try to mathematically describe the spread of an epidemic in a population \cite{Book_2008,Book_2018}. The Susceptible-Exposed-Infectious-Removed, SEIR, class of models are some of the simplest and most studied ones.
As an epidemic spreads, at any time $t$, let $I(t)$ be the number of infected individuals and $S(t)$ be the number of susceptible individuals in the population. Furthermore, let $R(t)$ be the number of individuals removed from the epidemic dynamics (i.e. the number of those who have either died till time $t$ or who have recovered till this time). We assume that those individuals who have recovered from the disease can not become susceptible again.
Since the functions $I(t), S(t)$ and $R(t)$ give the number of individuals, their codomain must be the set of non-negative integers and hence, strictly speaking, these functions can not be continuous.
Moreover, the true dynamics of the number of individuals during the spread of an epidemic is stochastic in nature.
We shall mostly deal with the fractions $s = S/N$, $i=I/N$, $r = R/N$, where $N$ is total population (which we assume changes only negligibly during the course of the spread of the epidemic). These quantities also change by discrete amounts but if the minimal possible change is sufficiently small, we can think of $s(t)$, $i(t)$ and $r(t)$ as differentiable functions.

In SEIR model \cite{Book_2008,Book_2018}
of the dynamics of epidemics, these quantities evolve in accordance with the following differential equations
 \begin{eqnarray}
 s'(t) & = & - \beta~ s(t)~  i(t) \; , \label{eq:s-smooth}\; \\
 e'(t) & = & \beta~ s(t)~  i(t) - \sigma~ e(t) \; ,  \label{eq:e-smooth}\; \\
 i'(t)  & = & \sigma~ e(t) - \gamma~ i(t) \; , \label{eq:i-smooth}\; \\
 r'(t) & = &  \gamma~ i(t) \label{eq:r-smooth}\; \; .
 \end{eqnarray}
Here, the parameter $\gamma$ is known as recovery rate (since $1/\gamma$ is the average duration of recovery or average infectious period) while the parameter $\sigma$ is known as incubation rate (since $1/\sigma$ is the mean incubation period). 
The values of these parameters, for COVID-19 have been experimentally estimated (see e.g. \cite{params} and \cite{jcm9040967}).

The parameter $\beta$ can be physically understood in the following manner:
in a homogeneously mixed population (in which everyone interacts with everyone else), if a randomly chosen susceptible individual experiences $\kappa$ contacts per unit time with other individuals, and if $c$ is the probability of disease transmission when this individual comes in contact with an infectious individual, then, the quantity $\beta$ is defined by the relation: $\beta \equiv - \kappa \ln (1-c)$ \cite{Book_2008}. For sufficiently small $c$, 
\begin{equation}
 \beta = \kappa c \; ,
\end{equation}
thus, $\beta$ is the product of contact rate and disease transmission probability and it itself is often simply called ``the contact rate."
Note that by its very definition, this quantity can not be negative.

Various mitigation measures will ensure that both $\kappa$ (the rate of contact) and $c$ (the probability of disease transmission) will change. It is this time dependence of the quantity $\beta$ which we wish to extract from the data.
Finally, for SEIR models (ignoring the changes in population due to births and deaths), the quantity  called ``Basic reproduction number," ${\cal R}_0$, which is
the expected number of cases directly generated by one case in a population where all individuals are susceptible to infection (however, see also e.g. \cite{ComplexityR0}), is given by
\begin{equation} \label{eq:R0}
{\cal R}_0 = \frac{\beta}{\gamma} \; .
\end{equation}
This can be used to define an effective time dependent reproduction number ${\cal R} (t)$ by the same expression. Given the effective time dependence of $\beta$, ${\cal R} (t)$ defined in this manner can be easily determined.

\subsection{Theory and observational data}
\label{sec:theory_obs}

Let us now connect the theoretical description of the spread of the pandemic to observations.
Let $j$ be an index characterising the day number, then, the observational data about quantities such as 
$\mathscr{C}_j$ (the number of confirmed cases of the pandemic till the day characterised by the index $j$), 
$\mathscr{D}_j$ (the number of people who have died till the day characterised by the index $j$) and $\mathscr{R}_j$ (the number of people who have recovered till the day characterised by the index $j$), is available. 

Note how $\mathscr{C}_j$ differs from a discretised version of $I(t)$. Since $\mathscr{C}_j$
represents the number of individuals infected till day $j$, it is a cumulative quantity, on the other hand, $I(t)$ is the number of infected individuals at time $t$.
Given the definitions of these quantities, one expects that
\begin{equation} \label{eq:icdr}
I_{j} = \mathscr{C}_j - \bigg[ \mathscr{D}_j  + \mathscr{R}_j  \bigg] \; .
\end{equation}
This can be used to find the number of people infected on day $j$ i.e. $I_j$ from the observed data.
In the next subsection, we will see how this can be used to reconstruct an effective time dependent $\beta$ parameter.

In the very beginning, when the very first cases of a pandemic are observed, the cumulative number of dead and the cumulative number of recovered are both zero and at that stage $I_j$ and $\mathscr{C}_j$ are identical. 
As the cumulative number of dead and recovered increases, the difference between $I$ and $\mathscr{C}$ also increases, but as long as the cumulative number of dead and the cumulative number of recovered are small compared to the number of confirmed cases, $I_{j}$ remains close to $\mathscr{C}_j$. 
At a much later stage of the evolution of the pandemic, $I_{j}$ begins to decrease, while $\mathscr{C}_j$, being a  cumulative quantity, never decreases.

Finally, let us note that, in the discrete form, the SEIR evolution equations become (for time dependent $\beta$),
 \begin{eqnarray}
 s_{j+1} - s_{j} & = & - \beta_j ~ s_j~  i_j \; ,  \label{eq:s} \\
 e_{j+1} - e_{j} & = & \beta_j~ s_j~  i_j - \sigma~ e_j \; , \label{eq:e} \\
 i_{j+1} - i_{j}  & = & \sigma~ e_j - \gamma~ i_j \; , \label{eq:i} \\
 r_{j+1} - r_{j} & = &  \gamma~ i_j \; . \label{eq:r}
 \end{eqnarray}

\section{Strategy for reconstruction}
\label{sec:strategy}

\subsection{Reconstructing the effective contact rate}

Given the time series data of number of infected individuals each day, $I$, we follow the following simple steps to reconstruct, not only $e_n$, $r_n$ and $s_n$, but also the effective time dependent contact rate $\beta_n$.
The method described here can be applied to any population which is sufficiently large, but, to illustrate the method, we shall apply it to countries. Here are the steps of the procedure:

\begin{enumerate}
 \item For any country of interest, we obtain the time series data of $\mathscr{C}_j, \mathscr{D}_j, \mathscr{R}_j$ from \cite{data_ref} and obtain $I_j$ by following the procedure described in the last section.
 We focus attention to the data for only those countries for which the number of tests per million, at the time of writing, is sufficiently large (at least a few thousand, see table \ref{table:multicountries}).
We shall also see what happens when we do work with the data for a country for which the number of tests per million is very low. 
 \item We only begin to use the data from the time when the cumulative number of confirmed cases is greater than 25, this corresponds to the day which we shall characterise by the index $n=1$. Let $t_i$ be the day before this. Similarly, let us assume that the time series data is available till time $t_f$, which also corresponds to $n = n_{\rm max}$.
 \item For each country, we note down the date on which lockdowns began (time $t_l$) as well as its population ($N$).
We can now obtain the values of $i_n$ by dividing the number of infected by the population.
 \item We use the values of the recovery rate $\gamma$ and the incubation rate $\sigma$ available in the literature \cite{params}, \cite{jcm9040967}.
 \item At this stage, we use eq (\ref{eq:i}) to find out $e_n$ for all $n$ except $n = n_{\rm max}$ (since we will not know $i_{n_{\rm max}+1}$).
 \item Now we use Eq (\ref{eq:r}) to find out $r_n$ for all $n$. In order to do so, we shall need $r_1$. To find this, we assume that, for the day which corresponds to  $n=1$, the number of removed individuals is equal to the sum of number of  recovered individuals and the number of dead individuals. We shall eventually see what happens when we relax this assumption.   
 \item Knowing $i_n$ and having found $e_n$, $r_n$, we could find $s_n = 1 - (i_n + e_n + r_n)$. Since we do not know $e_n$ for $n = n_{\rm max}$, we would only know $s_n$ for $n < n_{\rm max}$.
 \item Now, use Eq (\ref{eq:e}) to find $\beta_n$
 \begin{equation} \label{eq:beta}
  \beta_n = \frac{e_{n+1} - e_n + \sigma e_n}{s_n ~ i_n} \; .
 \end{equation}
Since $e_n$ is only known till $n_{\rm max} - 1$, the maximum value of $n$ for which we can find $e_{n+1}$ in the RHS of the above equation must be $n_{\rm max} - 1$. Thus, we can find $\beta_n$ for $1\le n \le n_{\rm max} - 2$. 
\end{enumerate}

Before proceeding, we note that, from Eq (\ref{eq:e}), Eq (\ref{eq:i}) and Eq (\ref{eq:beta}), it is easy to show that:
\begin{equation} \label{eq:beta_full}
 \beta_n = \frac{1}{s_n i_n \sigma} \bigg[ i_{n+2} + (\gamma + \sigma - 2)~i_{n+1} + (\sigma - 1) (\gamma - 1)~i_n \bigg] \; .
\end{equation}
Moreover, from Eq (\ref{eq:R0}), it is clear that this procedure will also give us a time dependent effective reproduction number, ${\cal R}$ which is defined by Eq (\ref{eq:R0}) but for a time dependent $\beta$.
As a test of self-consistency of this procedure, one could check whether Eq (\ref{eq:s}) gets satisfied, we confirm that this is indeed the case.
The above steps let us reconstruct the effective, time dependent $\beta$ using the SEIR evolution equations. This kind of information about the effective $\beta$ caused by various mitigation measures is of utmost importance if we wish to be able to simulate the mitigation measures: this can in fact predict the long term spread of the epidemic. In the rest of this subsection, we shall understand various issues associated with this reconstruction.

\subsubsection{An example of reconstruction}

\begin{figure}
  \includegraphics[width = 0.45\textwidth]{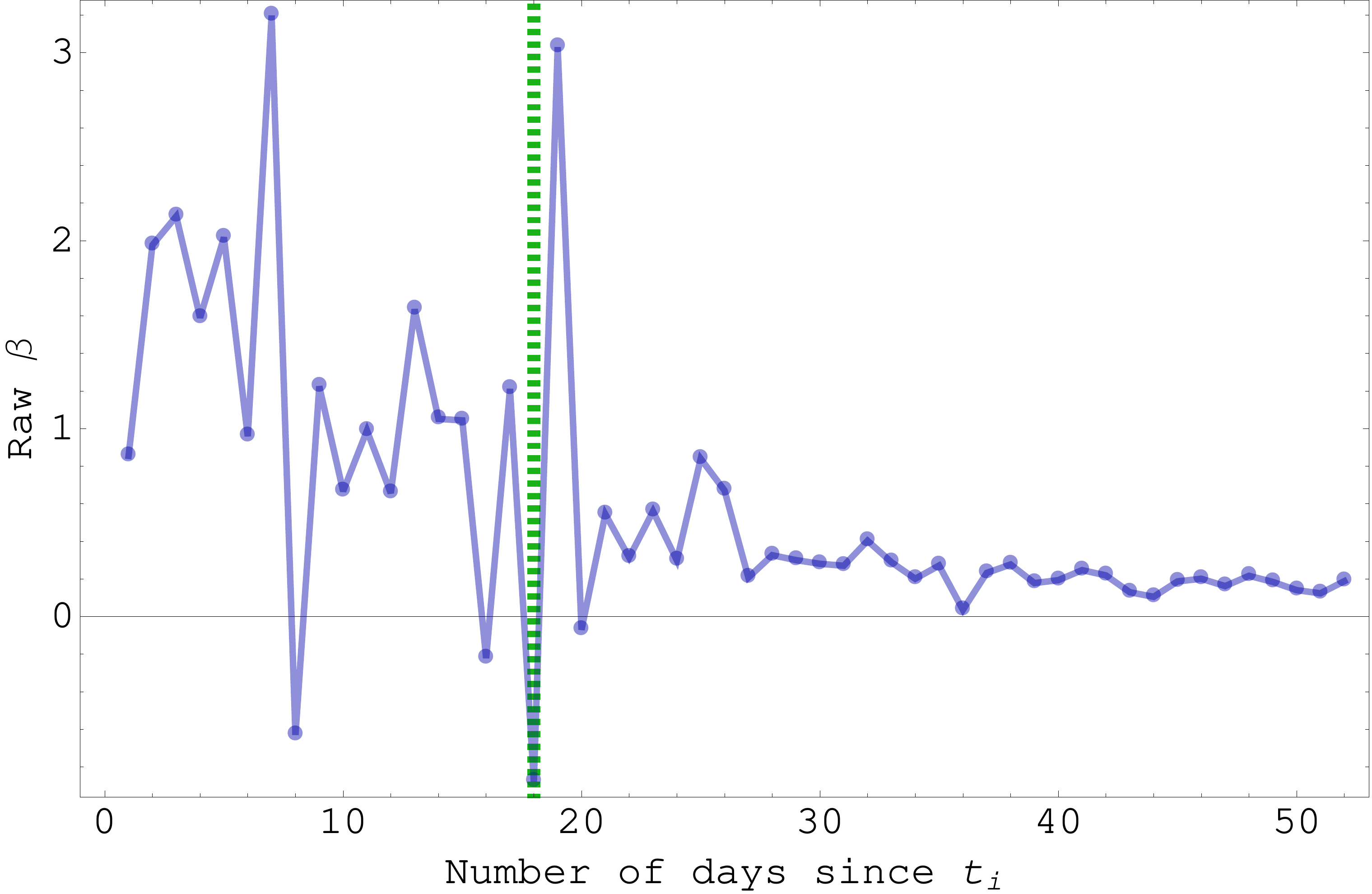}
  \caption
 {Reconstructed raw $\beta$ for Italy (blue data points and blue lines connecting them).  
 The vertical straight line is the day after nationwide lockdowns began.
 For this reconstruction, we chose $1/\gamma = 7$ days and $1/\sigma = 7$ days.
 }
  \label{beta-italy}
\end{figure}

Let us look at the results of this procedure for a specific example case. We first show these results for Italy, for which, at the time of writing, there have been 18,481 tests per million (i.e. $1.85 \%$) of the population, so that one can be quite confident in trusting the data. For Italy, the population (as of 2019) is $60.4$ million and, on Feb 22, 2020, the number of infected became greater than 25, i.e. the day before this date corresponds to the time value $t_i$ (and Feb 22, 2020 itself corresponds to $n=1$). Similarly, for Italy, lockdown time $t_l$ corresponds to the date March 09, 2020, i.e. $n_l = 17$. Finally, if we analyse the Italy data till April 15, 2020, we would have $n_{\rm max} = 54$.
In fig (\ref{beta-italy}), we show the result of step 8 above on the time series data for Italy. 
It is easy to see that the behaviour of this reconstructed $\beta$ (which we call raw $\beta$ in the figure) changes after the lockdown, in fact, the average value of reconstructed $\beta$ goes down after lockdown. The average value of $\beta$ before lockdown is 1.2, while the average value of $\beta$ after lockdown is only 0.3.

\subsubsection{Remarks about occasional negative values} \label{sec:negative}

Looking at Fig (\ref{beta-italy}), it is clear that this reconstruction procedure can occasionally give negative values of $\beta_n$ for certain values of $n$: this is obviously unphysical. From Eq (\ref{eq:beta}), it is clear that this happens whenever  
$e_{n+1} < (1-\sigma) e_n$. 
\footnote{A similar condition for the existence of negative values of $\beta_n$ in terms of $i_n$ can also be found from Eq (\ref{eq:beta_full}).}
In the early stage of growth of the epidemic, we do not expect that $e_{n+1}$ shall be less than $e_n$. Thus, the occasional negative values can only arise due to fluctuations. 
Before proceeding, we must note that, at a much later stage, we do expect that $e_{n+1}$ will be (typically) less than $e_{n}$, but this also does not correspond to negative $\beta$, it just means that in the RHS of Eq (\ref{eq:e-smooth}), the second term is dominant.

Since the description in terms of smooth functions is a coarse grained description of true stochastic dynamics, 
the actual evolution of $i_n$ is not determined by deterministic differential equations (or even difference equations), it is a stochastic quantity whose evolution is only roughly captured by these equations. Thus, when we use Eq (\ref{eq:s} -\ref{eq:r}) to determine $\beta_n$ from the data which in reality is determined not by these equations, but by some stochastic dynamics, we expected that due to the fluctuations, inferred $\beta_n$ will occasionally be negative.  
Thus, it is expected that only a smoothed out form of $\beta$ obtained by reconstruction can be a sensible quantity.

\subsubsection{The effect of changing $r_1$}

In step 6 of the reconstruction procedure, we had chosen value of $r_1$. One expects that, since the variable $r_n$ does not impact the dynamics of any other variable, this choice should have no impact. On the other hand, the value of $r_n$ does determine the value of $s_n$ which gets used in finding $\beta$. It is thus important to ask ourselves whether this has any impact on the reconstructed $\beta$. We repeated the analysis for different chosen values of $r_1$ and found no change in the reconstructed $\beta$. 

\subsubsection{The effect of changing $\gamma$ and $\sigma$}

\begin{figure*}[htp]
  \centering
  \hspace*{-1cm} 
 \subfigure[ $~$Reconstructed $\beta$ as we change $\gamma$ (for $\sigma = 1/9~{\rm day}^{-1}$).]{\includegraphics[width = .5\textwidth,keepaspectratio]{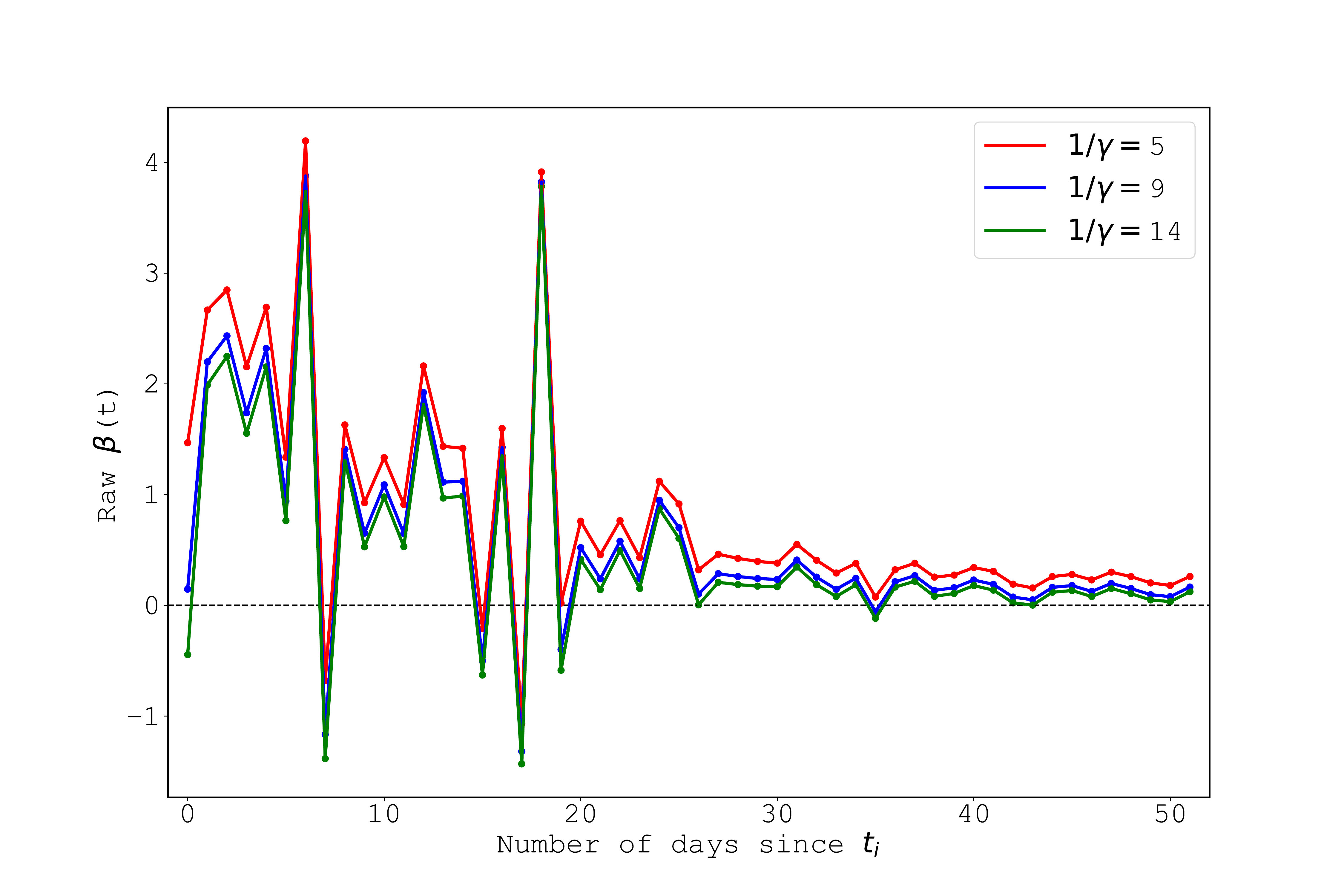}} \quad
  \subfigure[ $~$Reconstructed $\beta$ as we change $\sigma$ (for $\gamma  = 1/9~{\rm day}^{-1}$).]{\includegraphics[width = .5\textwidth,keepaspectratio]{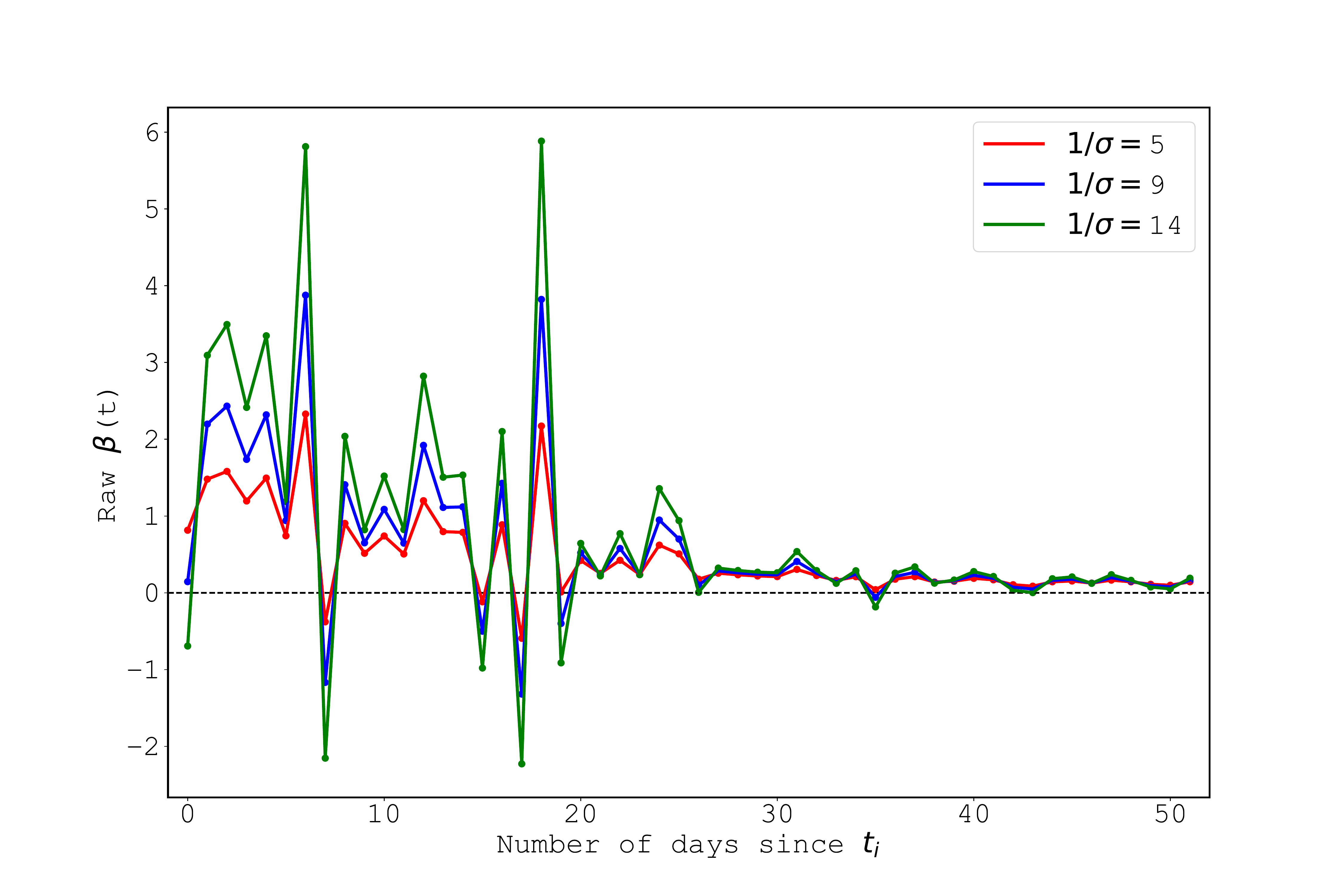}}  
  \caption{
The uncertainties in reconstruction of $\beta$ due to experimental uncertainties in the known values of the recovery rate $\gamma$ and the incubation rate $\sigma$.
It is worth noting that decreasing $\gamma$ leads to a slight decrease in reconstructed raw $\beta$ while decreasing $\sigma$ leads to a fairly substantial increase in reconstructed $\beta$. 
  }
   \label{effect-change}
\end{figure*}

Till now, we have looked at the reconstructed $\beta$ for a fixed set of values of the recovery rate ($\gamma$) and incubation rate ($\sigma$). One might wonder to what extent this reconstruction depends on these values. In fig (\ref{effect-change}), we show how the reconstructed $\beta$ changes for various choices of $\gamma$ and for various choices of $\sigma$. It is clear that the experimental uncertainties in the values of recovery rate and incubation rate lead to uncertainties in our reconstructed $\beta$. In particular,
decreasing the central value of $\gamma$ leads to a slight decrease in reconstructed raw $\beta$ while decreasing the central value of $\sigma$ leads to a substantial increase in reconstructed $\beta$. This is of course easy to understand from Eq (\ref{eq:beta_full}): 
in units of ${\rm day}^{-1}$, $\gamma$ and $\sigma$ are small compared to $1$, so, in the approximation in which we completely ignore them, $\beta_n \propto 1/\sigma$: this explains the strong inverse proportionality of $\beta$ to $\sigma$. In order to find the dependence on $\gamma$, we can no longer ignore $\gamma$ as compared to 1, so, there will be a weak dependence on $\gamma$.

As the experimental limits on the recovery rate and the incubation rate improve, our knowledge of $\beta(t)$ will also improve.
We also note in passing that the instantaneous effective reproduction number, $\mathscr{R}$ will scale inversely as we change $\gamma$.

\subsection{Post-processing the raw reconstruction}
\label{sec:postprocess}

We saw in \textsection \ref{sec:negative}, that spurious fluctuations and occasional unphysical negative values turn up in reconstructed $\beta$ because of random statistical fluctuations in the input time series data of infected fraction of the population.
We expect the actual $\beta$ to be a positive function of time, and we would like to use a more smooth functional form of $\beta$ (so that it can be described by a few numbers). 
I.e., the reconstruction procedure described above gives us a raw contact rate $\beta_n$ and we need to find a smoothed version of $\beta$ which one can easily work with.

\begin{figure*}[hbtp]
  \centering
  \hspace*{-1cm} 
 \subfigure[ $~$Two moving averages of raw $\beta$.]{\includegraphics[width = .5\textwidth,keepaspectratio]{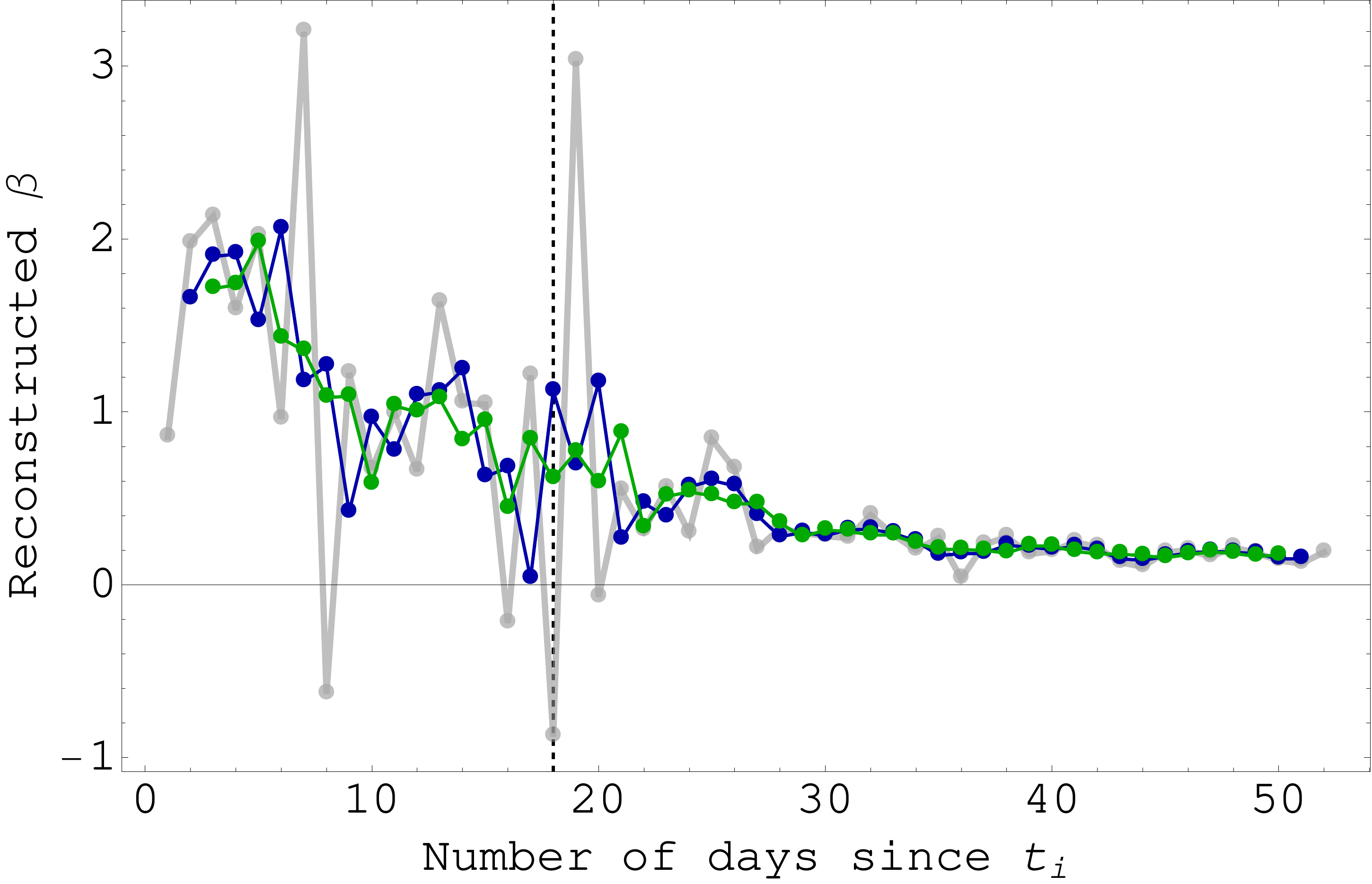}} \quad
  \subfigure[$~$Fitting two smooth functions to raw $\beta$.]{\includegraphics[width = .5\textwidth,keepaspectratio]{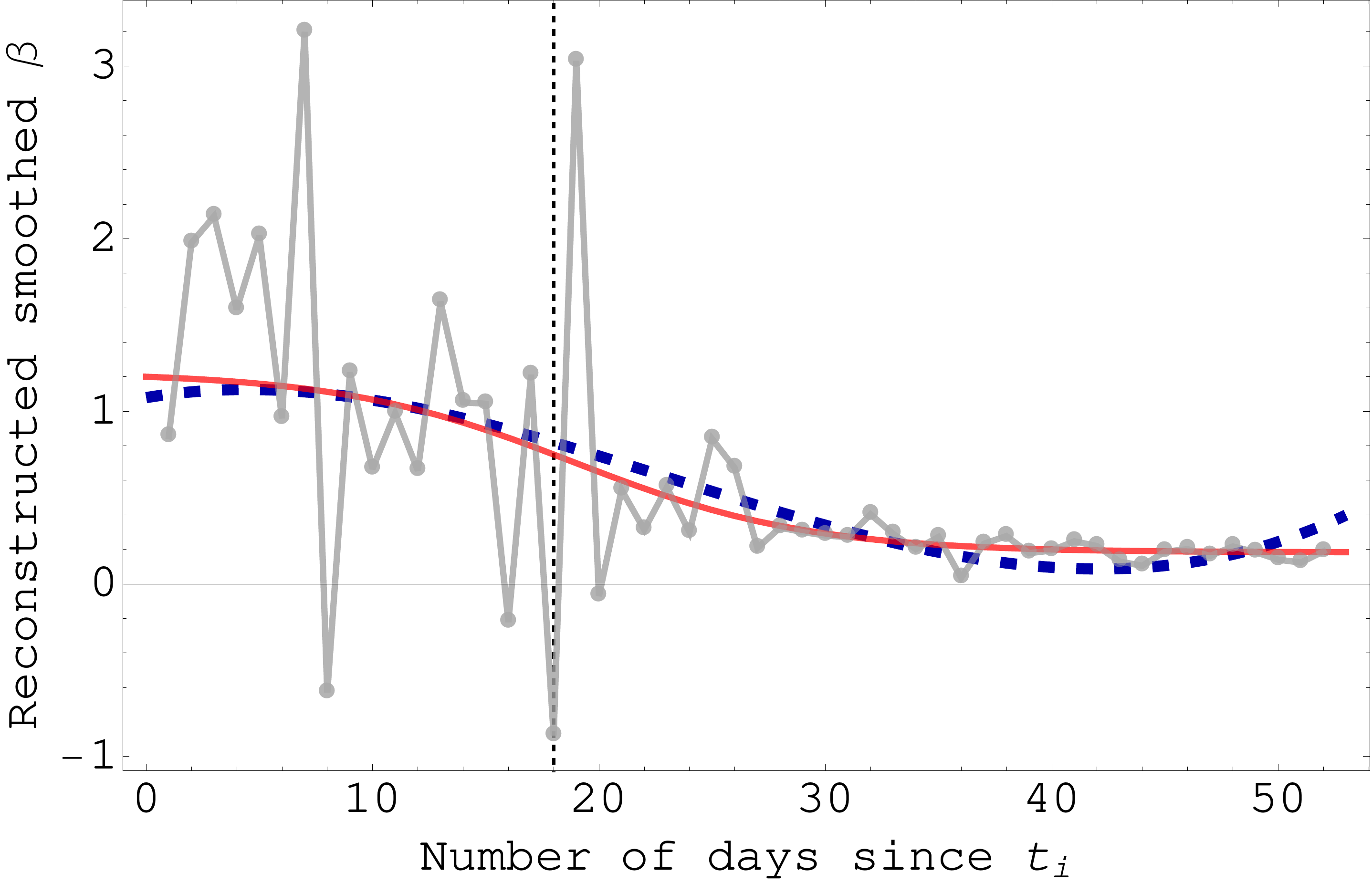}}  
  \caption{
Some of the ways of post-processing the raw contact rate to obtain a smoothed contact rate 
(the grey points in both the plots correspond to the raw $\beta$).
{\bf Left:} 
The blue points correspond to a moving average of three days duration while the green points correspond to the moving average of five days duration. 
{\bf Right:}
The dashed blue curve in this figure is a smooth, third order polynomial fit to raw $\beta$ (given by Eq (\ref{eq:poly})) and the solid red curve is a fit to the ``step-down function" (defined by Eq (\ref{eq:step})).
}
   \label{smooth}
\end{figure*}

To check that smoothing gets rid of occasional negative entries, we found the moving average of reconstructed raw $\beta$ for a window function of duration three days and five days. The results of this are shown in the left plot of fig (\ref{smooth}).
As expected, moving averages of longer duration help us in getting rid of not only the occasional negative values of $\beta$, but also, spurious fluctuations. In the left plot of fig (\ref{smooth}), one should also note that as expected, the data for three-days moving average starts one day after $n=1$ and ends one day before $n=n_{\rm max}$. Similarly, the data for five-days moving average starts two days after $n=1$ and ends two days before $n=n_{\rm max}$.

In addition, to describe $\beta$ by a smooth function, so that it can be specified by only a few numbers,
we fitted (a) a low degree polynomial to the reconstructed raw $\beta$, and,
(b) a ``step-down function", of the form 
\begin{equation} \label{eq:step}
 \beta_{\rm step} (t) = A \tanh \left(\frac{t - t_{\rm off} }{t_w} \right) + B \; ,
\end{equation}
to the reconstructed raw $\beta$. The parameters of this fitting function turn out to be 
$A= -0.52, B = 0.70, t_{\rm off} = 18.90, t_{w} =  10.35$ while
a cubic polynomial fit to the $\beta$ turns out to be
\begin{eqnarray} \label{eq:poly}
 \beta_{\rm poly}(t) &=& 1.08 ~+~ 0.02 ~t ~-~ 2.72 \times 10^{-3} ~t^2 \nonumber \\
 &+& 3.90 \times 10^{-5} ~t^3 \; .
\end{eqnarray}
Both of these functions are shown in the right plot of  fig (\ref{smooth}). 
The parameters of the step down function are especially noteworthy: $t_{\rm off} = 18.9$, in comparison, $t_l$ (the lockdown day) in this case is $17$, also, since $t_w = 10.35$, this needs to be compared with the infectious period (which was set to be $7$ days for this run).

As we shall argue in the next section, all of these smoothed out forms of $\beta$ are as good as the raw $\beta$ in modelling the effects of various mitigation measures.
Any of the post-processed i.e. smoothed $\beta$ found here can then be used to model the actual spread of a pandemic such as COVID-19. 

\subsection{Simple applications}

\subsubsection{Using the form of $\beta(t)$ to find long term evolution}

\begin{figure}
  \includegraphics[width = 0.45\textwidth]{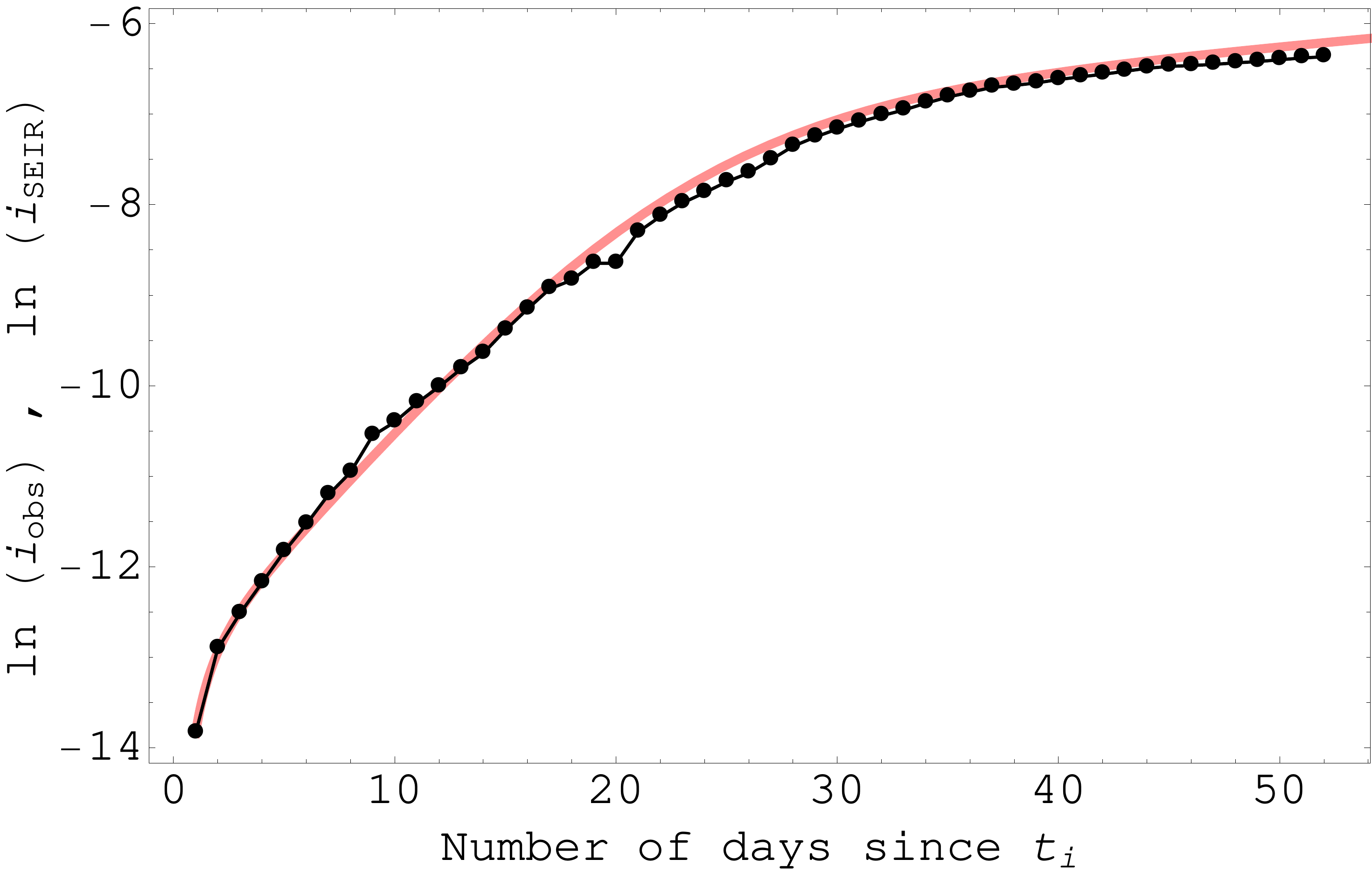}
  \caption
 {
  In this fig, on the vertical axis, we have natural logarithm of the fraction of infected population $i(t)$. The red solid curve corresponds to $i(t)$ found by evolving Eqs (\ref{eq:s-smooth}-\ref{eq:r-smooth}) with a time dependent smoothed out $\beta$ approximated by the ``step-down function" (defined by Eq (\ref{eq:step})). The data points correspond to $i(t)$ obtained from observations. Despite replacing the raw reconstructed $\beta$ by the smooth step-down $\beta$, the agreement between the two is noteworthy.
  }
  \label{seir_step}
\end{figure}

One way to ensure that the procedure described in the last section is self-consistent, is, to evolve SEIR equations i.e. Eqs (\ref{eq:s-smooth}-\ref{eq:r-smooth}), for the time dependent $\beta$ obtained in the last section. 
In order to do so, we need to set the values of various parameters and initial conditions in Eqs (\ref{eq:s-smooth}-\ref{eq:r-smooth}).
Needless to say, for $\gamma$ and $\sigma$, we need to choose the same values for which we performed the reconstruction. 
For initial condition for $i(t)$, we use the observed value of $i (t_i)$ obtained using Eq (\ref{eq:icdr}). 
On the other hand, for the fraction of population which is removed, we set $r(t_i) = r_1$, while for the fraction of population exposed, we use the value of $e(t_i)$ obtained by our reconstruction procedure. Next we need to use the reconstructed $\beta$ in SEIR evolution equations. In order to do so, we can either work with raw $\beta$, or, we can work with one of the smoothed forms of $\beta$ found in the last section. 
In fig (\ref{seir_step}), we show $i(t)$ found out after using the ``step-down" form of $\beta (t)$ (defined in Eq (\ref{eq:step})) as well as the $i(t)$ obtained from observations. From the fact that the two curves are so close, we learn that instead of working with raw $\beta$, we could also work with the ``step-down" form of $\beta$ in order to find the long term evolution of the infected population. 
Similarly, in modelling the effects of various mitigation measures, other smoothed out forms of $\beta$ (such as the moving average or the polynomial fit) are almost as good as raw $\beta$ itself.

At this stage it is easy to see that one could use Eqs (\ref{eq:s-smooth}-\ref{eq:r-smooth}), to find the location of the peak, the extent of flattening of the curve (due to mitigation measures) and other long term effects if the same mitigation measures continue. Furthermore, if in the near future, the mitigation measures are relaxed, and we wish to reintroduce them in a few weeks time, the effective instantaneous contact rate found by the procedure described in the last section, can be used to predict how the epidemic will spread in the next wave.

\subsubsection{Applying to other populations}

In the last section, since replacing raw $\beta_n$ by a smoothed $\beta(t)$ did not have significant impact on the evolution of the fraction of infected population (see fig (\ref{seir_step})), one could simply work with one of the smoothed forms of $\beta(t)$. 
Since a smoothed form of $\beta(t)$ can be described by only a handful of numbers (such as $A$, $B$, $t_{\rm off}$ and $t_w$ for the step down function or the coefficients of various powers of $t$ in the (low degree) polynomial function), we can say that
these few numbers carry information about the entire history of mitigation measures for Italy during the period of spread of the pandemic.
Thus, depending on the population of interest (such as a chosen country or a chosen city), 
and depending on the chosen parameterisation of $\beta(t)$, we can find the few numbers which specify the mitigation history.
For the spread of COVID-19 pandemic, the social interventions taken up by various countries are very well known and well-recorded (in the form of public policies, media reports etc). Thus, we can relate the mitigation measures taken up to the few numbers we use to parameterise $\beta(t)$. 

\begin{figure*}
  \includegraphics[width = \textwidth,keepaspectratio]{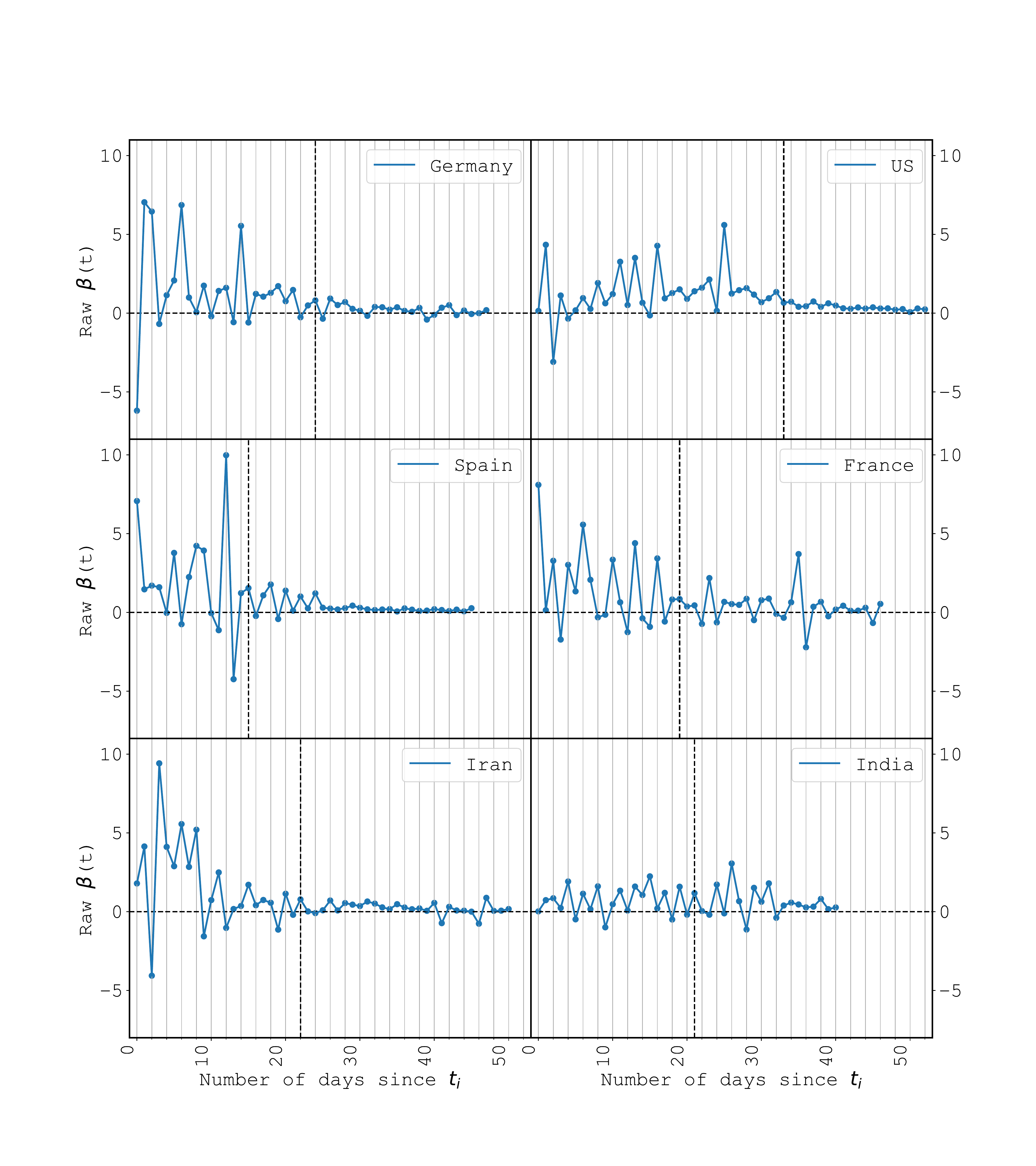}
  \caption
 {
  Reconstructed raw $\beta$ from the time series data for six other countries. 
    Note that time $t_i$ corresponds to different dates for different countries (see Table \ref{table:multicountries} for details).
  The vertical dashed line corresponds to $t_l$, the time at which lockdown is implemented. The dramatic change in the behaviour of effective $\beta$ after lockdown is quite apparent for all countries except India for which, the number of tests per million is lowest (among the countries considered here), see table \ref{table:multicountries}.
  }
  \label{beta_multiple}
\end{figure*}

Finally, it is interesting to ask how the effective instantaneous contact rates of various countries compare with each other. 
For each country, one could either work with the raw data for $\beta_n$ or with one of the smoothed (descriptions described in the last section). We provide the raw $\beta_n$ for six additional countries in fig (\ref{beta_multiple}) and some useful basic information about these countries is given in table \ref{table:multicountries}. The parameters $\gamma$ and $\sigma$
are both set to be $\frac{1}{7}~{\rm day}^{-1}$. 
Note that the date on which lockdowns begin are different for different countries, thus, the duration between the date of lockdown and the date till which data is used in this work, is different for different countries.
Note also that the extent of lockdowns is different for these different countries.
In fig (\ref{beta_multiple}), the impact of lockdowns can be seem from the behaviour of raw $\beta_n$ for all countries except India. This is because the number of tests per million for India is very low, see table \ref{table:multicountries}. 
Since many of the plots in fig (\ref{beta_multiple}) have several negative entries, one would need to employ smoothing procedures to get a smooth, positive post-processed $\beta(t)$.

\begin{table}
\begin{tabular}
 {c l r r r r } 
\hline 
\hline
 &  &  &  &  &  \\ 
Sr.  & Country & Population & No. of tests & Starting  & Lockdown \\ 
No. &  & in millions & per million & date & date \\ 
 &  &  &  &  &  \\ 
\hline
\hline 
 &  &  &  &  &  \\ 
1 & Germany & 83.7 & 15,730 & Feb 26 & Mar 21  \\ 
 &  &  &  &  &  \\ 
2 & U.S.A. & 330.6 & 8,156 & Feb 21 & Mar 25  \\ 
 &  &  &  &  &  \\ 
3 & Spain & 46.7 & 7,593 & Feb 28 & Mar 14  \\ 
 &  &  &  &  &  \\ 
4 & France & 65.2 & 5,114 & Feb 27 & Mar 17  \\ 
 &  &  &  &  &  \\ 
5 & Iran & 83.7 & 3,136 & Feb 22 & Mar 15  \\ 
 &  &  &  &  &  \\ 
6 & India & 1377.1 & $< 200$ & Mar 04 & Mar 25  \\ 
 &  &  &  &  &  \\ 
7  & Italy & 60.4 & 19,935 & Feb 22 & Mar 09 \\ 
 &  &  &  &  &  \\ 
\hline 
\end{tabular}
\caption{
This table contains some relevant information about the six countries for which the results are shown in Fig (\ref{beta_multiple}) and Italy (for which we have illustrated most of our formalism (the dates are for the year 2020). Thee starting date is the date on which the number of infected became greater than 25. Data about various countries has been obtained from \cite{data_ref} and \cite{lockdown_ref}.
\label{table:multicountries}
}
\end{table}

\section{Discussion}

In the absence of vaccines and treatment for a pandemic, several mitigation measures are taken to slow down its spread and to ensure that the healthcare system does not get overwhelmed. These measures include personal preventive measures (such as the use of face masks, self quarantine etc), social distancing measures (such as closing down of schools and cancellation of social gatherings etc), travel restrictions, curfews and so on.
When these drastic measures are practised on a large scale, they often lead to severe socioeconomic disruptions.
Given this, it is extremely important that we should be able to compare the economic costs of disruptions to the extent to which the spread of a pandemic can be arrested by these measures. 

This is a task which can potentially be handled by sufficiently well predictive mathematical models of the spread of pandemics. Needless to say, any reliable forecasting for a system as complex as a city or a country is a colossally difficult task. 
Often, one has to build a virtual copy of the entire population of interest in one's simulation, taking into account such diverse effects as migrations, number of contacts per day, the effect of testing, contact tracing etc. Despite this, the predictions of the model depend on the values of a large number of parameters whose values can only be roughly estimated at best.
Moreover, all of this modelling needs to be done while unprecedented mitigation measures are taking place. This just leads to the fact that, typically, models which try to take into account a very detailed description of the spread of the pandemic still prove to be inadequate.

While detailed modelling, which takes into account these various effects, can be done, in this paper, we assumed that, at a very coarse grained level of description, all of this amounts to causing the effective $\beta$ in SEIR models to become time dependent. If that is so, it is interesting to see how this effective $\beta$ changes in time because of various social interventions for various populations. Keeping this motivation in mind, we presented a procedure which can be used to reconstruct this time dependent (i.e. instantaneous) $\beta$ from the time series data of the number of confirmed cases. 

We began in section \textsection \ref{sec:reminder} by reminding the reader some fundamentals of SEIR models, particularly the physical significance of the quantity $\beta$.
In section \textsection \ref{sec:theory_obs}, we described how the variables turning up in the theoretical formulation are related to the observed quantities and illustrate it with an example. 
Then, in section \textsection \ref{sec:strategy}, we began by elaborating on the procedure for obtaining a raw form of $\beta$ by reconstructing from the time series data by making use of SEIR evolution equations. It turns out that this raw form of $\beta$ can occasionally take unphysical negative values. We have explained what causes this to happen. Furthermore, the reconstruction procedure depends on the values of other parameters such as recovery rate (denote by $\gamma$) and incubation rate (denoted by $\sigma$) whose experimentally known values have experimental uncertainties. Since this will lead to uncertainties in the determination of raw $\beta$, we found out how much this uncertainty in raw $\beta$ is.

Next, in section \textsection \ref{sec:postprocess}, we presented methods by which one can obtain smoothed form of $\beta$ free from spurious fluctuations and unphysical negative values. A smoothed $\beta$ can be described by a very few numbers, these few numbers carry information about the entire history of mitigation measures during the spread of the pandemic. Since for every country with known mitigation measures, these handful of numbers can be determined, we can find which mitigation measures lead to which behaviour of $\beta$.
We then illustrated how smoothed $\beta$ could be used to evolve SEIR equations. This can be used to find the evolution of the number of infected individuals if the mitigation measures are known beforehand.
Finally, we provided the raw $\beta$ for six countries to demonstrate the method.
Since many of the entries in raw $\beta_n$ turn out to be negative, one needs to apply smoothening procedures to obtain a physically acceptable $\beta(t)$ from raw $\beta_n$.

In summary, we have presented a method which can be used to extract a raw instantaneous effective contact rate for every population which undertakes mitigation measures. We described procedures for post-processing this raw contact rate to obtain a more physically acceptable instantaneous effective contact rate. This processed contact rate can be described by a few parameters. In future, one could find this instantaneous processed contact rate for various populations and relate it to the mitigation measures being put in place. Eventually, this can help us better understand to what extent a given mitigation measure affects the spread of the epidemic.

\begin{acknowledgements}
The work of M.D. is supported by Department of Science and Technology, Government of India under the Grant Agreement number IFA18-PH215 (INSPIRE Faculty Award).
\end{acknowledgements}


\begin{thebibliography}{99}

\bibitem{WHO}
Coronavirus disease (COVID-2019) situation reports: 
\href{https://www.who.int/emergencies/diseases/novel-coronavirus-2019/situation-reports}{https://www.who.int/}

\bibitem{Gorbalenya-2020}
Gorbalenya, A.E., Baker, S.C., Baric, R.S. et al. 
{\it The species Severe acute respiratory syndrome-related coronavirus: 
classifying 2019-nCoV and naming it SARS-CoV-2}. \href{https://doi.org/10.1038/s41564-020-0695-z}
{Nat Microbiol 5, 536?544 (2020)}.
\bibitem{Wu-2020}
Wu, F., Zhao, S., Yu, B. et al. {\it A new coronavirus associated with human respiratory disease in China}. 
\href{https://doi.org/10.1038/s41586-020-2008-3}{Nature 579, 265?269 (2020)}. 
\bibitem{Zhou-2020}
Zhou, P., Yang, X., Wang, X. et al.
{\it A pneumonia outbreak associated with a new coronavirus of probable bat origin } . 
\href{https://doi.org/10.1038/s41586-020-2012-7}{Nature 579, 270?273 (2020)}. 



\bibitem{Ferguson-2006}
Ferguson, N., Cummings, D., Fraser, C. et al. 
{\it Strategies for mitigating an influenza pandemic}. 
\href{https://doi.org/10.1038/nature04795}{Nature 442, 448?452 (2006)}.  

\bibitem{Huaiyu-2020}
Huaiyu Tian et. al., 
{\it An investigation of transmission control measures during the first 50 days of the 
COVID-19 epidemic in China}, 
\href{https://science.sciencemag.org/content/early/2020/03/30/science.abb6105}{Science  31 Mar 2020}


\bibitem{Prem-2020}
Kiesha Prem et. al., 
{\it The effect of control strategies to reduce social mixing on outcomes of the COVID-19 epidemic in Wuhan, China: a modelling study}, \href{https://doi.org/10.1016/S2468-2667(20)30073-6}
{Lancet. 2020; 395: 689-697} 
\bibitem{Singh-2020}
Rajesh Singh, R. Adhikari, {\it Age-structured impact of social distancing on the COVID-19 epidemic in India},
\href{https://arxiv.org/abs/2003.12055}{arXiv:2003.12055 [q-bio.PE]}
\bibitem{time-dependent-beta}
Elena Loli Piccolomini, Fabiana Zama,
{\it Preliminary analysis of COVID-19 spread in Italy with an adaptive SEIRD model}
\href{https://arxiv.org/abs/2003.09909}{https://arxiv.org/abs/2003.09909}
\bibitem{Imperial-college}
Walker, Patrick GT; Whittaker, Charles; Watson, Oliver et al.
{\it The Global Impact of COVID-19 and Strategies for Mitigation and Suppression}
\href{https://www.imperial.ac.uk/media/imperial-college/medicine/sph/ide/gida-fellowships/Imperial-College-COVID19-Global-Impact-26-03-2020v2.pdf}{Imperial College COVID-19 Response Team}

\bibitem{Berger}
Berger, David W and Herkenhoff, Kyle F and Mongey, Simon,
{\it An SEIR Infectious Disease Model with Testing and Conditional Quarantine}
\href{http://www.nber.org/papers/w26901}{National Bureau of Economic Research, Working Paper Series, 26901, 2020.}

\bibitem{jia-2020}
J.~Jia et al., {\it Modeling the Control of COVID-19: Impact of Policy Interventions and Meteorological Factors},  \href{https://arxiv.org/abs/2003.02985}{(arXiv:2003.02985[q-bio.PE]}

\bibitem{pandey-2020}
G.~Pandey et. al., {\it SEIR and Regression Model based COVID-19 outbreak predictions in India}, \href{https://arxiv.org/abs/2004.00958} {(arXiv:2004.00958[q-bio.PE]}

\bibitem{Pribylova-2020}
L.~Pribylova and Veronika Hajnova, {\it SEIAR model with asymptomatic cohort and consequences to efficiency of quarantine government measures in COVID-19 epidemic}, \href{https://arxiv.org/abs/2004.02601} {(arXiv:2004.02601[q-bio.PE]}

\bibitem{Das-2020}
S.~Das et. al., {\it Critical community size for COVID-19 -- a model based approach to provide a rationale behind the lockdown}, \href{https://arxiv.org/abs/2004.03126}, {(arXiv:2004.03126[q-bio.PE]}

\bibitem{Castilho-2020}
C.~Castilho et. al., {\it Assessing the Efficiency of Different Control Strategies for the Coronavirus (COVID-19) Epidemic}, \href{https://arxiv.org/abs/2004.03539}, {(arXiv:2004.03539[q-bio.PE]}

\bibitem{Sardar-2020}
T.~Sardar, S.~Nadim, J.~Chattopadhyay, {\it Assessment of 21 Days Lockdown Effect in Some States and Overall India: A Predictive Mathematical Study on COVID-19 Outbreak}, \href{https://arxiv.org/abs/2004.03487}, {(arXiv:2004.03487[q-bio.PE]}



\bibitem{Book_2008}
Keeling, Matt J. and Rohani, Pejman,
{\it Modeling Infectious Diseases in Humans and Animals},
Princeton University Press, 2008.



\bibitem{Book_2018}
Li, Michael Y.
{\it An Introduction to Mathematical Modeling of Infectious Diseases}
Springer International Publishing, Mathematics of Planet Earth, Vol 2, 2018.

\bibitem{ComplexityR0}
Delamater, Paul L et al.,
{\it Complexity of the Basic Reproduction Number (R0).}
Emerging infectious diseases vol. 25,1 (2019): 1-4. doi:10.3201/eid2501.171901


\bibitem{params}
Li, Qun et al.,
{\it Early Transmission Dynamics in Wuhan, China, of Novel Coronavirus - €"Infected Pneumonia}
New England Journal of Medicine, 382, 13, 1199-1207, 2020
\href{https://doi.org/10.1056/NEJMoa2001316}{https://doi.org/10.1056/NEJMoa2001316}

\bibitem{jcm9040967}
Minah Park et. al.,
{\it A Systematic Review of COVID-19 Epidemiology Based on Current Evidence},
Journal of Clinical Medicine, 9, 4, 967 (2020), 
\href{https://www.mdpi.com/2077-0383/9/4/967}{https://www.mdpi.com/2077-0383/9/4/967}


\bibitem{data_ref}
E. Dong, H. Du, and L. Gardner. 
{\it An interactive web-based dashboard to track covid-19 in real time} The Lancet Infectious Diseases, 2020. URL: \href{https://linkinghub.elsevier.com/retrieve/pii/S1473309920301201}{link}. See also,
The online interactive dashboard hosted by the Center for Systems Science and Engineering (CSSE) at Johns Hopkins University, Baltimore, MD, USA,
\href{https://github.com/CSSEGISandData/COVID-19/tree/master/csse_covid_19_data}{link}


\bibitem{lockdown_ref}
COVID-19 Lockdown dates by country: A list of countries and the dates that each country went into lockdown.
\href{https://www.kaggle.com/jcyzag/covid19-lockdown-dates-by-country/version/3}{link}







\end{thebibliography}
\end{document}